\documentstyle[12pt,dmtrieste,psfig]{article}

\def\mincir{\raise -2.truept\hbox{\rlap{\hbox{$\sim$}}\raise5.truept
\hbox{$<$}\ }}
\def\magcir{\raise -2.truept\hbox{\rlap{\hbox{$\sim$}}\raise5.truept
\hbox{$>$}\ }}
\def\rf{\par\noindent\hangindent 20pt}
\catcode`\@=11
\long\def\@makefntext#1{
\protect\noindent \hbox to 3.2pt {\hskip-.9pt
$^{{\ninerm\@thefnmark}}$\hfil}#1\hfill}                

\def\@makefnmark{\hbox to 0pt{$^{\@thefnmark}$\hss}}  

\def\ps@myheadings{\let\@mkboth\@gobbletwo
\def\@oddhead{\hbox{}
\rightmark\hfil\ninerm\thepage}
\def\@oddfoot{}\def\@evenhead{\ninerm\thepage\hfil
\leftmark\hbox{}}\def\@evenfoot{}
\def\sectionmark##1{}\def\subsectionmark##1{}}

\begin{document}

\centerline{\normalsize\bf 
A Physical Model for Baryons in Clusters of Galaxies}

\vspace*{0.6cm}
\centerline{\footnotesize Paolo Tozzi}
\baselineskip=13pt
\centerline{\footnotesize\it II Universit\`a di Roma}
\baselineskip=12pt
\centerline{\footnotesize\it }
\centerline{\footnotesize E-mail tozzi@roma2.infn.it}
\vspace*{0.3cm}
\centerline{\footnotesize Alfonso Cavaliere}
\baselineskip=13pt
\centerline{\footnotesize\it II Universit\`a di Roma}
\baselineskip=12pt
\centerline{\footnotesize\it }
\centerline{\footnotesize E-mail cavaliere@roma2.infn.it}
\centerline{\footnotesize and}
\vspace*{0.3cm}
\centerline{\footnotesize Nicola Menci}
\baselineskip=13pt
\centerline{\footnotesize\it OAR}
\centerline{\footnotesize E-mail menci@coma.mporzio.astro.it}

\vspace*{0.9cm}
\abstracts{
The X-ray emission from clusters of galaxies is one of the best
observational probe to investigate the distribution of dark matter 
at intermediate and high redshifts.  Since the disposition of the intracluster
plasma (ICP) responsible of the emission is crucial to link X-ray properties
to the global properties of the dark matter halos, we propose a 
semi--analytical approach for the diffuse baryons.  
This comprises the following blocks: Monte Carlo ``merging histories'' to 
describe the dynamics of dark matter halos; the central {\sl hydrostatic} 
disposition for the ICP; conditions of shock, or of closely adiabatic 
compression at the 
{\sl boundary} with the external gas, preheated by stellar energy feedbacks. 
From our model we predict the $L-T$ correlation, consistent with the data as 
for shape and scatter.  }

\normalsize\baselineskip=15pt
\section{Introduction}

Groups and clusters of galaxies constitute cosmic structures sufficiently 
close to equilibrium and with sufficient density contrast ($\delta\approx 
2\, 10^2$ inside the virial radius $R$) as to yield definite 
observables.  
They are dominated by dark matter (hereafter DM), while the baryon fraction
is observed to be less than $20$\%.  The great majority of these baryons
are in the form of {\sl diffuse plasma} (ICP) with densities 
$n\sim 10^{-3}$ cm$^{-3}$ and virial temperatures $k\,T\sim GM\,m_H/10 R\sim 
5$ keV, and are responsible for powerful X--ray luminosities $L\sim 10^{44}$ 
erg/s by optically thin thermal bremsstrahlung.  As the plasma is a good 
tracer of the potential wells, much better than member galaxies, the X--ray 
emission is a powerful tool to investigate the mass distribution out to 
moderate and high redshifts.  The ICP temperature directly probes the height 
of the potential well, with the baryons in the  role of  mere tracers; 
on the other hand, the luminosity, with its strong dependence on density  
($L\propto n^2$), reliably probes the baryonic content and distribution.  
Statistically, a definite $L$--$T$ correlation is observed (albeit with 
considerable scatter), and this provides the crucial link to relate the X-ray 
luminosity functions with the statistics of the dark mass $M$ or with that of 
the corresponding $T$.  

Many numerical experiments, using hydrodynamical N--body (see especially 
Gheller, Pantano \& Moscardini 1997, Bryan \& Norman 
1997), provide a comprehensive tool to model the ICP distribution in the 
potential wells.  However, such numerical experiments still do 
not have enough dynamic range 
to describe DM and ICP over the full range from $\sim 50$ Mpc associated with 
the large scale structures (which guide the ongoing mergers of DM halos), 
to the inner $50$ kpc where the ICP yields a substantial contribution to $L$, 
and moreover do not include properly non--gravitational effects, which
instead cannot be ignored.  

The simplest semi--analytical approach is constituted by the Self Similar 
model (Kaiser 1986) which include only 
gravity and assumes the ICP amount to be proportional to the DM at all $z$ 
and $M$.  This leads to a relation $L\propto T^2$ conflicting with the 
observed correlation for rich clusters, which is close to $L\propto T^3$ 
(David et al. 1993; Mushotzky \& Scharf 1997).  A missing ingredient 
is thought to be the stellar energy feedback by supernovae.  

The above features motivate us to develop a comprehensive semi--analytical 
model for the ICP which include the effects of stellar energy scales.  First 
we assume that such energy input is efficient in depleting the
potential wells of the clusters progenitors, at $z\simeq 1 \div 2$, and in
pre-heating the intergalactic medium to temperatures in the range $T_1=0.1
\div 0.8$ keV as recently evidentiated in the outer cluster atmosphere
(Henriksen \& White 1996).  Then we describe clusters evolution as a sequence 
of hierarchical merging episodes of the DM halos, associated 
in the ICP to shocks of various strengths (depending on the mass ratio of the 
merging clumps), which provide the boundary conditions for the ICP to 
re--adjust to a new hydrostatic equilibrium.  In \S 3 we shall show that our 
predictions for the X-ray properties of clusters are consistent with recent 
data (see also Cavaliere, Menci \& Tozzi 1997, hereafter CMT97). 

\section{A physical model for the ICP}

The mass growth of dark halos leading to the formation of groups and
clusters, develop through the accretion of diffuse matter or massive
merging events between virialized halos.  In our approach 
the history of such episodes is followed in the framework of the 
hierarchical clustering by Monte Carlo simulations.  
During the mass growth, the pre--heated ICP is recovered through shocks 
of variable intensity, depending on the temperature ratio between the 
accreted gas and the virialized plasma contained in the {\sl main progenitor}.  
The ICP is reset to a {\it new} equilibrium after each episode of accretion 
or merging.  

\begin{figure}
\centerline{\psfig{figure=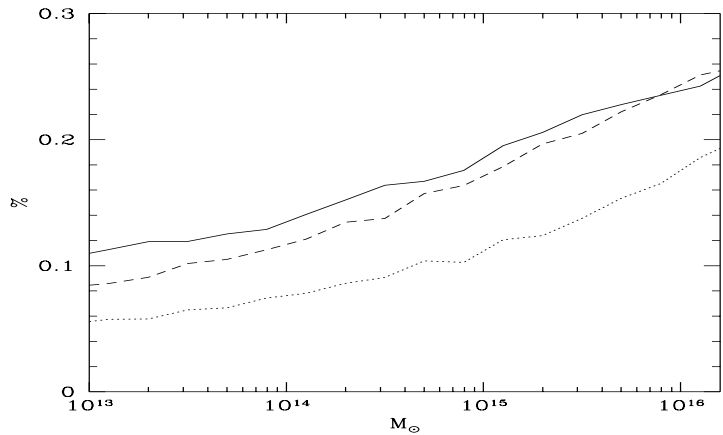,height=6truecm,width=10truecm}}
\caption{\sl Fraction of groups and clusters which have experienced at least 
one massive merging event with a mass ratio less then $2.5$ in the last
$2$ Gyrs.  Solid line: tilted CDM, dotted line: Open CDM, dashed line:
$\Lambda$CDM (see N. Menci this volume).  }
\label{disturb}
\end{figure}

In this framework it is possible to check the 
reliability of the assumption of equilibrium.  Roettiger, Stone \& Mushotzky 
(1997) show that after a major merging event, i.e., 
with a mass ratio less than $2.5$, the non thermal contribution to the pressure 
lasts less than two Gyrs.  Adopting this as a conservative rule to identify 
{\sl disturbed} objects, we can compute the fraction of clusters at 
redshift $z=0$ for which the hydrodynamical equilibrium does not fully apply.  
As is shown in fig. \ref{disturb} this fraction is less than 
$20$\% at cluster scales in most FRW universes.  Of course the situation is 
better for small objects, as they are in average older than rich clusters.

\subsection{Hydrodynamical equilibrium}

To compute the disposition of the ICP in equilibrium in a DM potential well, 
we have just to solve the hydrostatic equilibrium equation
\begin{equation}
{1\over {n}}{{dP}\over {dr}}=-G {{M(<r)}\over{r^2}}\, ,
\end{equation}
where $n(r)$ is the gas density profile, $M(<r)$ is the mass contained within 
the radius $r$ (dominated by DM) and $P$ is the pressure.  
To solve the equation we need the boundary condition, i.e. the density
of the gas at the virial radius $n(R)$, and the state equation, that we
write for a polytropic gas:
\begin{equation}
P(r)={{kT(r)}\over{\mu m_H}} {n^\gamma (r)}\, ,
\end{equation}
where $m_H$ is the proton mass, $\mu\sim 0.6$ is the mean molecular weight of 
the ICP, and $\gamma$, ranging from $1$ to $5/3$, is the polytropic index 
treated as a free parameter.  The resulting profile is: 
\begin{equation}
n(r)=n(R)\Big[ 1+ \beta \Big( {{\gamma -1}\over{\gamma}}\Big)
\Big( \phi(R)-\phi(r)\Big)\Big]^{1/(\gamma -1)}\, ,
\label{prof}
\end{equation}
where $\phi\equiv V/\sigma_r$ is the adimensional gravitational potential 
and $\beta\equiv \mu m_H \sigma_r/kT_2$, with $\sigma_r$ the line--of--sight
velocity dispersion of the DM particles.  Here $T_2$ is the temperature at
the virial radius.  The above equation can be considered a generalized 
$\beta$--model (Cavaliere \& Fusco--Femiano 1978), which reconduces 
to the usual isothermal case for $\gamma\rightarrow 1 $.  Before computing 
the ICP distribution, we then need the boundary condition $n(R)$.  

\subsection{Physics of shocks}

The key boundary condition is provided by the dynamic stress balance 
$P_2=P_1+m_H\,n_1\,v_1^2$, relating the exterior and interior 
pressures $P_2$ and $P_1$ to the inflow 
velocity $v_1$ driven by the gravitational potential at the boundary. 
Here $n_1$ is the baryon density external to the virial radius, and it
is assumed to be unbiased respect to the universal value, i.e. $n_1=\Omega_B 
\rho$ where $\rho$ is the total matter density.  We expect the inflowing 
gas to become supersonic close to $R$, when $m_H\,v_1^2> 2 kT_1$.  
In fact, many hydrodynamical simulations of loose gas accretion 
into a cluster (from Perrenod 1980 to Takizawa \& Mineshige 1997) show 
shocks to form, to convert most of the bulk energy into thermal energy, 
and to expand slowly remaining close to the virial radius for some dynamical 
times.  So we take $R$ as the shock position, and focus on nearly static 
conditions inside, with the internal bulk velocity $v_2<< v_1 $.  

The post-shock state is set by conservations across the shock 
not only of the stresses, but also of mass  and momentum, as described by the  
Rankine-Hugoniot conditions (see Landau \& Lifshitz 1959).  
These provide at the boundary the temperature jump $T_2/T_1$, and the
corresponding density jump $g\equiv n(R)/n_1$ which  reads 
\begin{equation}
g\Big({T_2\over T_1}\Big) = 
2\,\Big(1-{T_1\over T_2}\Big)+\Big[4\, 
\Big(1-{T_1\over T_2}\Big)^2 + {T_1\over T_2}\Big]^{1/2}
\label{gt}
\end{equation}
for a plasma with three degrees of freedom.  Eq. \ref{gt} includes 
both {\it weak} 
(with $T_2 \approx T_1$, appropriate for small groups accreting preheated gas, 
or for rich clusters accreting comparable clumps), 
and {\it strong} shocks (appropriate to ''cold inflow" as in rich 
clusters accreting small clumps and diffuse gas).  
From clusters to groups, the density jump $g(T)$ lowers from the maximum
value of $4$ towards unity.  

Given the inflow velocity $v_1$ and the nearly static post-shock condition 
$v_2<< v_1$, it is possible to work out the explicit expression
of the post-shock temperature $T_2$ in the form:
\begin{equation}
kT_2={{\mu m_H v_1^2}\over 3}\Big[ {{(1+\sqrt{1+\epsilon})^2}\over 4}
+ {7\over{10}}\epsilon -{{3}\over {20}}{{\epsilon^2}\over{(1+\sqrt{1+
\epsilon})^2}}\Big]\, ,
\end{equation}
where $\epsilon\equiv 15 kT_1/4 \mu m_H v_1^2$.  
For $\epsilon \gg 1$
the shock is weak and $T_2\simeq T_1$ is recovered as expected.  
In the case of strong shocks, $\epsilon \ll 1$ and the approximation 
$k\,T_2\approx -V(R)/3+3k\,T_1/2$ holds, where the second term is
the contribution from the non--gravitational energy input.  

\begin{figure}
\centerline{\psfig{figure=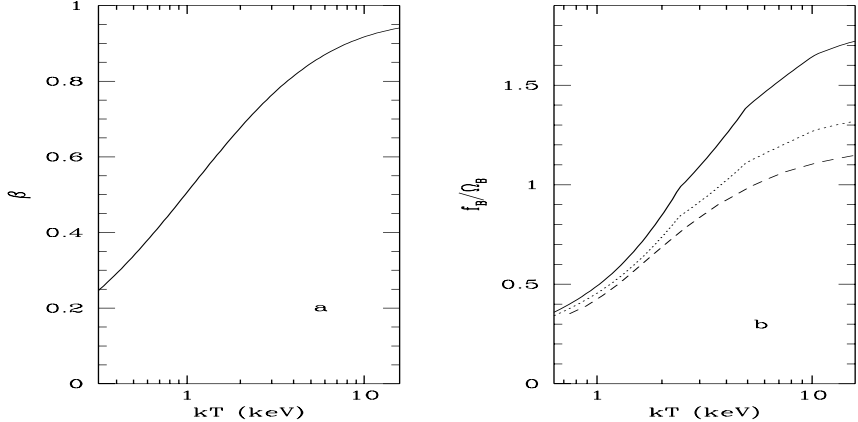,height=6.truecm,width=12truecm}}
\caption{\sl a) The $\beta(T)$ parameter entering equation \ref{prof}.  
b) Baryonic fraction respect to the universal value $\Omega_B$ for
different values of $\gamma$.  Continuous line: $\gamma =1$; dotted line:
$\gamma = 1.1$; dashed line: 
$\gamma = 1.2$.  }
\label{fig1}
\end{figure}

Now we can compute the ICP distribution, assuming a specific
choiche for the potential well (in the following we use forms given
by Navarro, Frenk \& White 1996, but of course this is not mandatory).  
First, we compute $\beta(T)$, which decline from $\mincir 1$ 
for rich clusters, 
to $\sim 0.4$ for poor groups, where the stellar competes with the 
gravitational energy, as shown in fig. \ref{fig1}a.  

\begin{figure}
\centerline{\psfig{figure=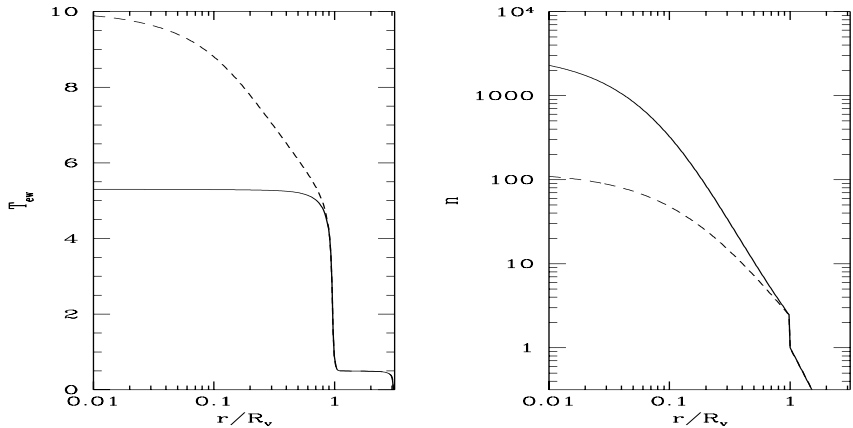,height=6truecm,width=12truecm}}
\caption{\sl Temperature and density profile for a cluster with $M=10^{15}
M_\odot$.  Continuous line: $\gamma =1$; dashed line: $\gamma = 1.2$.  }
\label{fig2}
\end{figure}

In fig. \ref{fig2}a,b we show the emission--weighted temperature and density
profile for two different values of $\gamma$ in the case of a rich
cluster with $M\sim 10^{15}M_\odot$.  A temperature profile with a mild 
decrease out to $r\sim$ 1 Mpc is in agreement  with the observations 
(Markevitch et al. 1997),  pointing toward a value $\gamma\mincir 1.2$.  
The corresponding baryonic fraction lowers down with the mass scale 
by a factor of three from clusters to groups (see fig. \ref{fig1}b).  

\section{The $L$-$T$ correlation}

The X-ray luminosity of a cluster with temperature profile $T(r)$ and density 
profile $n(r)$ can be written: 
\begin{equation}
L\propto \langle g^2(T) \rangle \int d^3 r\,{{n^2(r)}\over {n^2(R)}}\, 
T^{1/2}(r)\, \, .
\end{equation}

\begin{figure}
\centerline{\psfig{figure=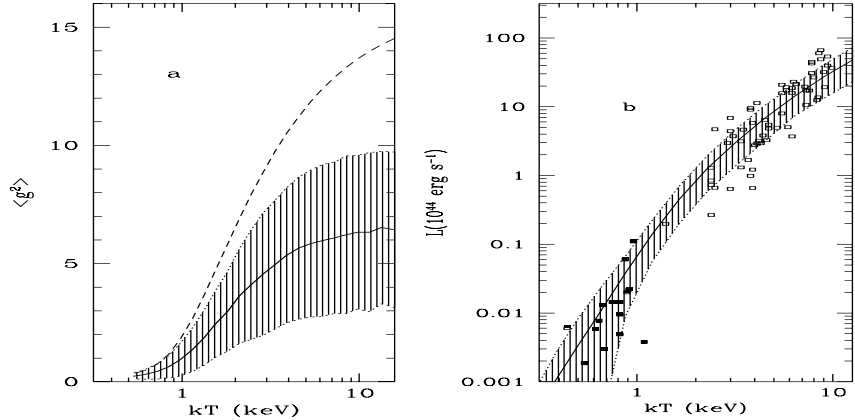,height=6truecm,width=12truecm}}
\caption{\sl a) The factor $\langle g^2 \rangle$ averaged over merging histories
(solid line) with $96$\% confidence level (shaded region).  The non--averaged
$g$ is plotted for comparison (dashed line).  b) $L$--$T$ relation compared 
with the data by David et al. (1993) for clusters (open squares) and 
Ponman et al. (1996) for groups (solid squares).  }
\label{fig4}
\end{figure}

In fact, before computing the $L$--$T$ relation, the statistical effect of the 
{\it merging histories} has to be taken into account.  For a cluster or group 
of a given mass (or temperature), the effective compression factor squared 
$\langle g^2 \rangle$ is obtained upon averaging eq. \ref{gt}
over the sequence of the DM 
merging events; in such events, $T_2$ is the virial temperature of 
the receiving structure, and $T_1$ is the higher  between the stellar 
preheating temperature and that from ``gravitational''
heating, i.e., the virial value prevailing in the clumps being accreted. 
All that is accounted for in our model using Monte Carlo simulations; these 
are based on merging trees corresponding to the excursion set approach of Bond 
et al. (1991), consistent with the Press \& Schechter (1974) statistics 
(see CMT97).  The averaged $\langle g^2 \rangle$ is lower than the 
$g^2$ computed with a single temparature $T_1$, 
because in many events the accreted gas is at a temperature higher than the 
preheating value.  In addition, an intrinsic variance 
is generated, reflecting the variance intrinsic to the merging histories
(see fig. \ref{fig4}a).  

The net result is shown in fig \ref{fig4}b.  In agreement with the 
observations (David et al. 1993; Ponman et al. 1996), 
the shape of the average $L-T$ relation flattens from $L\propto T^5$ at 
the group scale (where the nuclear energy from stellar preheating 
competes with the gravitational energy) to $L\propto T^3$ at the 
rich cluster scales. At larger temperatures the shape asymptotes to 
$L\propto T^2$, the self-similar scaling of pure gravity.  Notice the 
intrinsic scatter due to the variance in the dynamical merging histories, 
but amplified by the $n^2$ dependence of $L$.  We note that the shape of the 
$L$--$T$ relation is little affected by change of $\gamma$.  
The average normalization formally rises like $\rho^{1/2}(z)$, where $\rho$
is the effective external mass density which increases as $(1+z)^2$ 
(Cavaliere \& Menci 1997) in filamentary large scale structures hosting most 
groups and clusters.  This implies, e.g., an increase of $30$\% at $z=0.3$, 
consistent with the observations by Mushotzky \& Scharf (1997).  

\section{Conclusions}

The ICP {\it state} in the hierarchically evolving gravitational wells 
constitutes the focus of our new approach.  We propose that such state follows 
suit, passing through a sequence of equilibrium condition that we 
compute semi-analytically.  These computations comprise: 
the merging histories of the DM potential wells, obtained with a 
large statistics from Monte Carlo simulations of the hierarchical clustering; 
the inner {\sl hydrostatic} 
equilibrium disposition, updated after each merging episode; and the {\sl 
boundary} conditions provided by strong and weak shocks, or even by a closely 
adiabatic compression, depending on the ratio of the infall to the thermal 
energy in the preheated external medium.  

The results of our model depend on two parameters, the external temperature 
$T_1$ and density $n_1$, which are not free. Specifically, 
we use for $T_1$ the range $0.1\div 0.8 $ keV provided by the literature on 
stellar preheating.  The value of $n_1$ for rich clusters is related 
to the DM density by the universal baryonic fraction.  The expression of the 
bolometric luminosity is proportional to 
$g^2=(n_2/n_1)^2$, the square of the density jump at the 
bounding shock. The average of such factor over the merging histories coupled
with $\beta(T)$ is what gives to the statistical $L-T$ correlation the 
{\sl curved} shape shown in fig. \ref{fig4}b. 
In addition, our approach predicts an intrinsic variance of dynamical
origin due to the different merging histories, and built in the factor $g^2$.  
Moreover the decreasing temperature profiles are in agreement 
with the published data and with the results from advanced simulations. 
A straightforward application is the prediction for the X-ray statistics
in different FRW universes (see N. Menci, this volume, and Cavaliere, Menci \& 
Tozzi 1998).

\bigskip 
\bigskip

\section{References}
\bigskip


\rf{Bond, J.R., Cole, S., Efstathiou, G., \& Kaiser, N. 1991, ApJ
{\bf 379}, 440}

\rf{Bryan, G.L., \& Norman, M.L. 1997, preprint [astro-ph/9710107]}

\rf{Cavaliere, A, \& Fusco Femiano, R. 1976, A\&A, {\bf 49},137}


\rf{Cavaliere, A., Menci, N., 1997, ApJ, {\bf 480}, 132}

\rf{Cavaliere, A., Menci, N., Tozzi, P., 1997, ApJL {\bf 484}, 1 (CMT97)}

\rf{Cavaliere, A., Menci, N., Tozzi, P., 1998, ApJ submitted}

\rf{David, L.P., Slyz, A., Jones, C., Forman, W., Vrtilek, 
 S.D., \& Arnaud, K.A. 1993, ApJ, {\bf 412}, 479}

\rf{Edge, A.C., \& Stewart, G.C. 1991, MNRAS, {\bf 252}, 428}


\rf{Gheller, C., Pantano, O., \& Moscardini, L. 1997, preprint 
[astro-ph/9701170]}

\rf{Henriksen, M.J., \& White, R.E. 1996, ApJ, {\bf 465}, 515}


\rf{Kaiser, N. 1986, MNRAS, {\bf 222}, 323}

\rf{Landau, L.D., Lifshitz, E.M., 1959, {\it Fluid Mechanics}
(London, Pergamon Press), p. 329, 331}


\rf{Mushotzky, R.F., \& Scharf, C.A. 1997, ApJ, {\bf 482}, 13}

\rf{Navarro, J.F., Frenk, C.S., \& White, S.D.M. 1996, 
ApJ, {\bf462}, 563 }

\rf{Perrenod, S.C. 1980, ApJ, {\bf 236}, 373}

\rf{Ponman, T.J., Bourner, P.D.J., Ebeling, H., B\"ohringer, 
H. 1996, MNRAS, {\bf 283}, 690}

\rf{Press, W.H., \& Schechter, P. 1974, ApJ, {\bf 187}, 425}


\rf{Roettiger, K., Stone, J.M, \& Mushotzky, R.F 1997, preprint 
[astro-ph/9708043]}

\rf{Takizawa, M., \& Mineshige, S. 1997, preprint 
[astro/ph 9702047]}



\end{document}